\documentclass[prd,aps,nofootinbib,superscriptaddress]{revtex4}
\usepackage{epsfig}
\usepackage{amsmath, slashed}
\usepackage{xcolor}
\usepackage{appendix}
\usepackage{dsfont}
 \usepackage{mathrsfs}
\usepackage[normalem]{ulem}

\usepackage{bm}
\usepackage{amsfonts}

\allowdisplaybreaks
\begin{document}

\title{Baryon-antibaryon generalized distribution amplitudes and $e^+ e^-  \to B  \bar{B} \gamma$}

\author{Jing Han}
\affiliation{School of Physics, Zhengzhou University, Zhengzhou, Henan 450001, China}
\author{Bernard Pire }
\email{bernard.pire@polytechnique.edu}
\affiliation{CPHT, CNRS, \'Ecole polytechnique, Institut Polytechnique de Paris, 91128 Palaiseau, France}
\author{Qin-Tao Song}
\email[Contact author: ]{songqintao@zzu.edu.cn}
\affiliation{School of Physics, Zhengzhou University, Zhengzhou, Henan 450001, China}

\date{\today}

\begin{abstract}
{Baryon-antibaryon generalized distribution amplitudes (GDAs) give an access to timelike  gravitational form factors (GFFs) which are complementary to the spacelike ones which  can be deduced from the hadronic generalized parton distributions (GPDs) measured in deep exclusive electroproduction processes. They allow to probe the GFFs of unstable baryons in the baryon octet, since the second moments of hadronic  GDAs lead to the timelike GFFs.
These GDAs can be measured  in  the process $e^+ e^- \to B \bar{B}  \gamma$, in the generalized Bjorken regime where the invariant mass of the $B \bar{B}$ pair is near threshold  at high energy  facilities, such as BESIII, Belle II, and the  proposed  Super Tau-Charm Facility.
In this work, we investigate this process using the QCD collinear factorization framework, where the scattering amplitudes  are expressed in terms of the baryon timelike electromagnetic (EM) FFs and Compton FFs. We also provide a numerical estimate of the cross sections with a model for baryon-antibaryon GDAs.
Our work provides us a possibility to extract the timelike baryon GFFs from near future experimental measurements,
and these GFFs may be further used to study  longstanding questions in hadronic physics such as the baryon spin decomposition and D-term.}
\end{abstract}

\maketitle

\date{}

\section{Introduction}
\label{introduction}

The hadronic matrix elements of the energy-momentum tensor (EMT) are key physical quantities that reveal the inner structures of hadrons, and they are parameterized in terms of EMT  form factors (FFs), which are also known as gravitational FFs (GFFs) in the literature. The study of GFFs is related to currently hot topics in hadronic physics, such as spin decomposition~\cite{Ji:1996ek,Leader:2013jra,Ji:2020ena},
 mass radius~\cite{Polyakov:1998ze,Kumano:2017lhr,Freese:2019bhb,Wang:2023uek,GarciaMartin-Caro:2023toa,Wang:2024fjt}, and mechanical properties of the hadron~\cite{Polyakov:2018zvc,Burkert:2018bqq,Burkert:2023wzr,Lorce:2018egm,Kumericki:2019ddg,Li:2023izn,Nair:2024fit,Lorce:2025ayr,Broniowski:2025ctl}.
 It is impossible to directly probe the EMT FFs by experiment as the gravitational interaction is too week. Fortunately, the generalized parton distributions (GPDs) of the nucleon can be accessed by the exclusive processes such as deeply virtual Compton scattering (DVCS) and timelike Compton scattering (TCS), where only electromagnetic (EM) and strong  interactions are involved. The second moments of hadronic GPDs are the spacelike EMT FFs, and the study of GPDs is considered as an alternative to probe the GFFs. Thus, accessing hadronic GPDs with unprecedented precision is a top scientific priority for the forthcoming Electron-Ion Collider (EIC).

It is well known that most hadrons are not stable, and we cannot access their GPDs by  exclusive processes~\footnote{The $\pi$ meson case is an exception thanks to the study of Sullivan processes~\cite{Sullivan:1971kd, Amrath:2008vx, Chavez:2021koz, Castro:2025rpx}}.
For example, only the nucleon GPDs can be probed at existing or future lepton-nucleon facilities in the baryon octet and decuplet families. Transition GPDs like $N\to \Delta$ or $N \to N^*$ enter the description of electroproduction amplitudes such as DVCS with $N\to \Delta$ \cite{Kim:2024hhd} or $N \to N^*$ baryon transition (see the review \cite{Diehl:2024bmd} and references therein). They allow but a first step in the study of unstable baryons.

However, one can investigate the generalized distribution amplitudes (GDAs) of unstable hadrons through the production of a hadron-antihadron pair such as
$\gamma^{\ast} \gamma  \to  h \bar{h}  $~\cite{Kumano:2017lhr,Diehl:1998dk,Diehl:2000uv,Kivel:1999sd,Lorce:2022tiq,Lorce:2022cze,Guo:2025rhh,Song:2025zwl} and  $\gamma^{\ast} \to h \bar{h} \gamma$~\cite{Lu:2006ut, Pire:2023kng, Pire:2023ztb},  where the $h \bar{h}$ GDAs describe the amplitude of $q \bar{q} \to h \bar{h}$ under the QCD factorization.
The second moments of GDAs allow us to access  the timelike GFFs, and the spacelike  GFFs can be deduced from the timelike ones using  dispersion relation techniques. Thus, a complete picture of GFFs may be obtained from the study of the hadronic GDAs.
For the production of a pseudoscalar meson pair,
recent measurements of the cross sections for  $\gamma^{\ast} \gamma \to M \bar{M}$ and $\gamma^{\ast} \to M \bar{M} \gamma$ have been reported by the Belle~\cite{Belle:2015oin,Belle:2017xsz} and BaBar~\cite{BaBar:2015onb} collaborations, respectively. The dipion GDA and the pion GFFs were extracted from a theoretical analysis of Belle results~\cite{Kumano:2017lhr}.

We can extend the study of GDAs to the production of a baryon-antibaryon pair, and the amplitude of this process is expressed in terms of  baryon-antibaryon GDAs.
To satisfy the QCD factorization criteria~\cite{Diehl:1998dk}, the photon virtuality needs to be much larger than the invariant mass of the baryon pair.
The measurements of the reactions $\gamma^{\ast} \to B \bar{B} \gamma$ are possible at BESIII, Belle II, and the  proposed  Super Tau-Charm Facility (STCF)~\cite{Achasov:2023gey}, where the virtual photon comes from a $e^+e^-$ annihilation.
Since the real photon can be emitted from the initial lepton in $ e^+ e^- \to B \bar{B}  \gamma$, one also needs  to consider the contribution of the initial state radiation (ISR) process in the extraction of baryon GDAs; the ISR contribution is described by the baryon EM FFs, similarly to the Bethe-Heitler process in the DVCS reaction.
Recently, extensive experimental measurements have been reported on the timelike EM FFs of the baryon octet,
such as nucleons~\cite{CMD-3:2018kql, BESIII:2021tbq,BaBar:2013ves,BESIII:2021rqk,BESIII:2019hdp,BESIII:2022rrg,BESIII:2019tgo}, $\Lambda$~\cite{BESIII:2017hyw,BESIII:2019nep,BESIII:2023ioy}, $\Sigma$~\cite{BESIII:2020uqk,BESIII:2023ynq,BESIII:2023ldb,Belle:2022dvb,BESIII:2021rkn}, and $\Xi$~\cite{BESIII:2021aer,BESIII:2020ktn}. The  subsequent theoretical analysis of these measurements can be found in Refs.~\cite{Bianconi:2015owa,Tomasi-Gustafsson:2020vae,Lomon:2012pn,Qian:2022whn,Yan:2023nlb, Kuzmin:2024ozz, Bianconi:2022yjq, Yang:2024iuc, Cao:2021asd, Lin:2021xrc}.
Thus, the time has come to investigate the baryon-antibaryon GDAs in $ e^+ e^- \to B \bar{B}  \gamma$ with the baryon EM FFs as inputs.
Actually, the process $ e^+ e^- \to B \bar{B}  \gamma$ has frequently been used for the recent measurements of baryon EM FFs~\cite{BaBar:2013ves,BESIII:2023ioy,BESIII:2019tgo,Belle:2022dvb,BESIII:2023ldb}.
In this work, we investigate the baryon-antibaryon GDAs for the process $ e^+ e^- \to B \bar{B}  \gamma$ using QCD factorization, and theoretical formulas and numerical estimates are presented at leading twist and leading order in $\alpha_s$.

The present paper is organized as follows. In Sec.~\ref{kinematics}, we discuss the kinematics of $e^+e^-    \to B \bar{B}  \gamma$, and introduce the definition of the four leading-twist chiral-even \footnote{the four existing chiral-odd baryon-antibaryon GDAs do not contribute to our process.} baryon-antibaryon GDAs.
The relations between GDAs and timelike FFs are discussed in Sec.~\ref{ffs}.
The scattering amplitudes for the ISR and QCD processes are given, respectively,  in terms of timelike EM baryon FFs and Compton FFs in Sec.~\ref{crs}.
We introduce the models of EM FFs and baryon GDAs in Sec.~\ref{nucr}, and  derive numerical estimates  for the various contributions to the cross section  for the production of a proton-antiproton pair. In Sec.~\ref{Asymm} we calculate and discuss a forward-backward baryon asymmetry which is proportional to the interference of the ISR and QCD amplitudes, and is thus linear in baryon-antibaryon GDAs.
Our results are summarized in Sec.~\ref{summary}.

\section{Generalized distribution amplitudes in $e^+ e^- \to B \bar{B} \gamma$ }
\label{kinematics}

The baryon GDAs were first proposed to study the process $ \gamma \gamma \to B \bar{B}$ at large energy and momentum transfer~\cite{Diehl:2002yh} as well as the reversed process \cite{Freund:2002cq}, in a simplified handbag model;  these cross sections were expressed in terms of the first moments of GDAs.
In addition,  the meson GDAs can be probed in the process $\gamma^*  \to M \bar{M}  \gamma$, where virtual photon comes from  the $e^- e^+ $ annihilation, and its virtuality is large enough to satisfy the QCD factorization~\cite{Lu:2006ut}. In this work, we extend this process to the production of a baryon-antibaryon pair,
\begin{align}
e^-(k_1) e^+(k_2)  \to  \gamma^*(q_1) \to B(p_1) \bar{B}(p_2)  \gamma(q_2).
\label{eqn:baryon}
\end{align}
We define the following variables,
\begin{align}
s=(q_1)^2=(k_1+k_2)^2, \qquad u=(k_1-q_2)^2, \qquad  \hat{s}=P^2=(p_1+p_2)^2,  \qquad \Delta=p_2-p_1.
\label{eqn:kinva}
\end{align}
 The lightcone vectors $n$ and $\bar{n}$ are given by the photon momenta $q_1$ and $q_2$,
\begin{align}
n=\frac{\sqrt{2s}}{s-\hat{s}} q_2, \qquad \tilde{n}=\sqrt{\frac{2}{s}} (q_1-\frac{s}{s-\hat{s}}q_2).
\label{eqn:lcv}
\end{align}
For a Lorentz vector $v$, the lightcone components  $v^+=v \cdot n$ and  $v^-=v \cdot \bar{n}$ are defined, and the transverse component is given by
\begin{align}
v_T^{\alpha} =g_{T}^{\alpha \beta} v_{\beta}\,,
\label{eqn:dtrans}
\end{align}
with
\begin{align}
g_{T}^{\alpha \beta} =g^{\alpha \beta}-
n^{\alpha}\tilde{n}^{\beta}-n^{\beta}\tilde{n}^{\alpha}.
\label{eqn:gtmu}
\end{align}
The parameter $\zeta_0$  corresponds to the light-cone fraction of the momentum $\Delta$,
\begin{align}
\zeta_0= \frac{\Delta \cdot n}{P \cdot n}\,,
\label{eqn:skewness}
\end{align}
and one can obtain $\Delta_T^2=g_{T}^{\mu \nu} \Delta_{\mu} \Delta_{\nu}=4m^2-(1-\zeta_0^2)\hat{s}$, where $m$ is the baryon mass.

\begin{figure}[htp]
\centering
\includegraphics[width=0.5\textwidth]{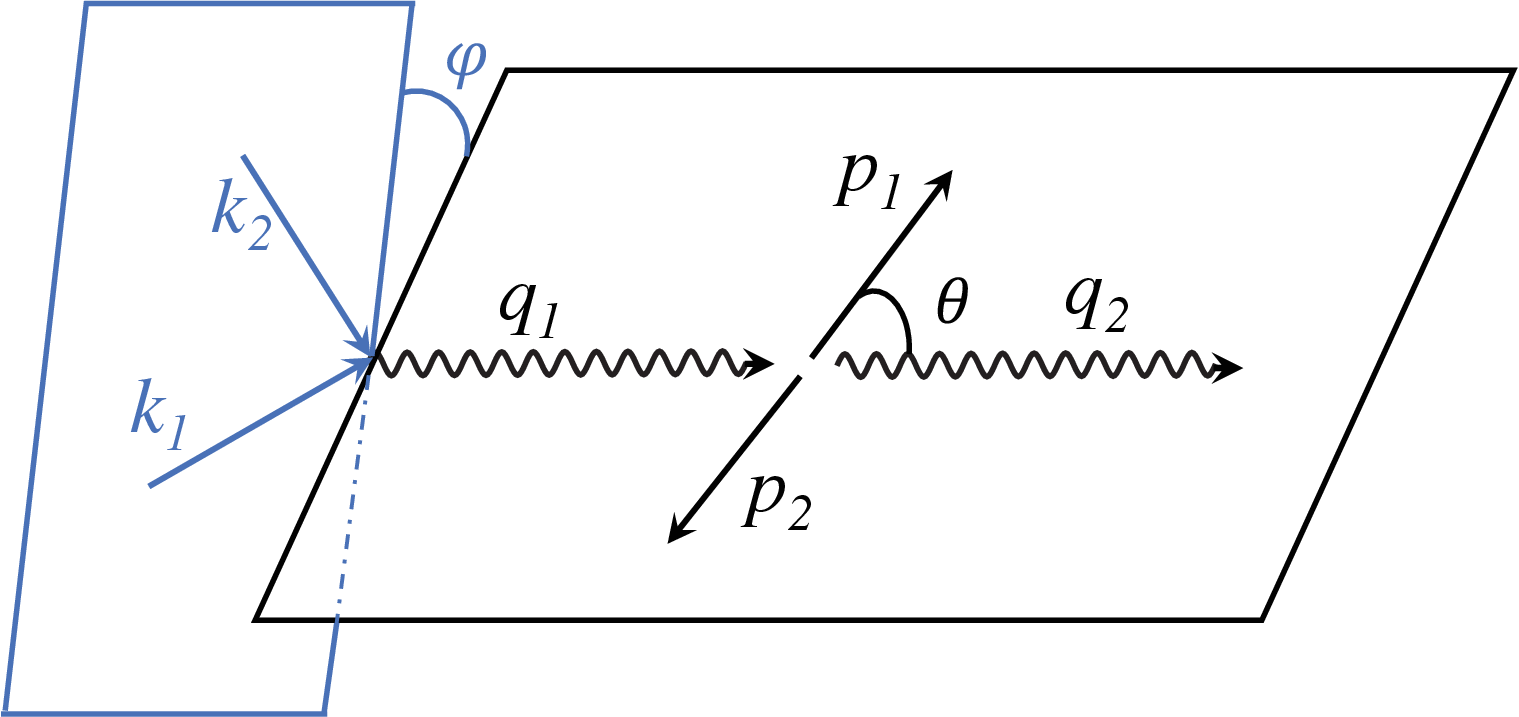}
\caption{
The Center-of-mass frame of the baryon-antibaryon pair for the process $ e^-(k_1)e^+(k_2)\to \gamma^{\ast}(q_1) \to  B(p_1) \bar{B}(p_2) \gamma(q_2)$.}
\label{fig:cmf}
\end{figure}
In the center-of-mass frame of the baryon - antibaryon pair, the $z$ axis is chosen along the momentum $q_1$, and the momenta of the baryon pair lie in the x-z plane, as illustrated by Fig.~\ref{fig:cmf}. The polar angle of $p_1$ is denoted as $\theta$,
\begin{align}
p_1=(p_1^0,\, |\bm{p_1}|\sin \theta,\, 0,\, |\bm{p_1}|\cos \theta)\,,
\label{eqn:p1m}
\end{align}
 and it can be expressed as
\begin{align}
\cos{\theta}=\frac{q_1\cdot(p_2-p_1)}{\beta_0\,(q_1\cdot q_2)}\,,
\label{eqn:pola}
\end{align}
with
\begin{align}
\beta_0=\sqrt{1-\frac{4 m^2}{\hat{s}}}.
\label{eqn:kinva1}
\end{align}
Combining  Eqs.~\eqref{eqn:skewness} and \eqref{eqn:p1m}, we obtain the relation $\zeta_0=\beta_0 \cos{\theta}$.
The azimuthal angle $\varphi$ between the lepton plane and the hadron plane is given by
\begin{align}
\sin{\varphi}=\frac{   -4  \epsilon_{\alpha \beta \gamma \delta}  q_1^{\alpha} q_2^{\beta}p_1^{\gamma} k_1^{\delta}  }{\beta_0 \sin{\theta} \sqrt{us \hat{s}(\hat{s}-u-s) }},
\label{eqn:amu}
\end{align}
where the convention $\epsilon^{0123}=1$ is used.

\begin{figure}[htp]
\centering
\includegraphics[width=0.4\textwidth]{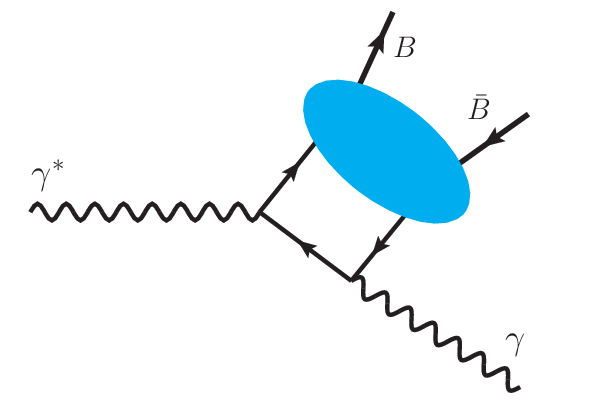}
\caption{
 The QCD subprocess (a)   $ \gamma^{\ast} \to  B \bar{B} \gamma$, where the blob represents the baryon-antibaryon GDAs. A  second  Feynman  diagram should be added, it is obtained by interchanging the photon vertices. }
\label{fig:gda}
\end{figure}

\begin{figure}[htp]
\centering
\includegraphics[width=0.5\textwidth]{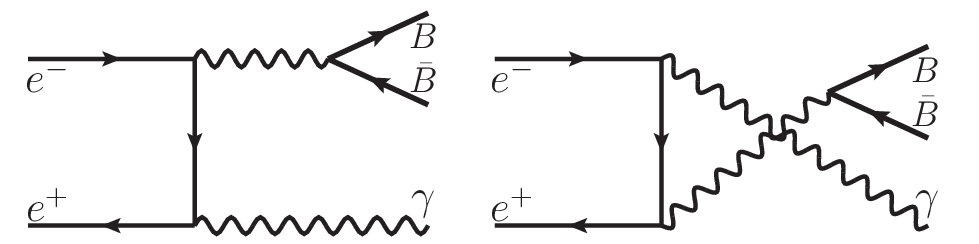}
\caption{
 The ISR subprocess (b)   $e^- e^+  \to \gamma^* \gamma \to B \bar{B}  \gamma$; the photon-baryon-antibaryon coupling is described by the timelike baryon EM FFs.}
\label{fig:codd}
\end{figure}

Two types of subprocesses contribute to the reaction  $e^- e^+  \to B \bar{B}  \gamma$, and we can distinguish them by the charge conjugation $C$ of the baryon-antibaryon pair.
The $C$-even baryon pair comes from the subprocess (a): $e^- e^+  \to \gamma^* \to B \bar{B}  \gamma$. At leading order in $\alpha_s$,  the amplitude of subprocess (a) can be factorized into
the hard part $\gamma^* \to q \bar{q}  \gamma$ and the soft part $q \bar{q} \to  B \bar{B}$ using the QCD factorization as illustrated by Fig.~\ref{fig:gda},  and the soft part is parameterized by  the baryon GDA.
In the lightcone gauge, we define the leading-twist chiral-even baryon-antibaryon quark GDAs for a spin-$\tfrac{1}{2}$ baryon pair~\cite{Diehl:2002yh},
\begin{equation}
\begin{aligned}
&P^+ \int dX^- \,e^{iz P^+ \cdot x^-}\, \langle \bar{B}(p_2) B(p_1)  | \,\bar{q}(-x^-) \gamma^+ q(0)\, | 0 \rangle
=
\Phi_V^q(z,\zeta_0, \hat{s}) \bar{u}(p_1) \gamma^+ v(p_2)+  \Phi_S^q(z,\zeta_0, \hat{s}) \frac{P^+}{2m} \bar{u}(p_1) v(p_2),\\
&P^+ \int dX^- \,e^{iz P^+ \cdot x^-}\, \langle \bar{B}(p_2) B(p_1)  | \,\bar{q}(-x^-) \gamma^+ \gamma_5 q(0)\, | 0 \rangle
=\Phi_A^q(z,\zeta_0, \hat{s}) \bar{u}(p_1) \gamma^+ \gamma_5 v(p_2)+  \Phi_P^q(z,\zeta_0, \hat{s}) \frac{P^+}{2m} \bar{u}(p_1) \gamma_5 v(p_2),
\label{eqn:gdame}
\end{aligned}
\end{equation}
where $z$ is the momentum fraction carried by the quark $q$, and we omit to write the  factorization scale dependence of the GDAs. One can obtain the following relations using the charge conjugation symmetry,
\begin{equation}
\begin{aligned}
 \Phi_i^q(z,\zeta_0, \hat{s})=&  \Phi_i^q(\bar{z},-\zeta_0, \hat{s}), \, \text{for} \, \, i=V, \, A, \, P, \\
 \Phi_S^q(z,\zeta_0, \hat{s})=& - \Phi_S^q(\bar{z},-\zeta_0, \hat{s}),
\label{eqn:chsy}
\end{aligned}
\end{equation}
and we note $\bar{x} \equiv 1-x$. In Fig.\ref{fig:codd}, the $C$-odd baryon-antibaryon pair is produced from the  ISR  subprocess (b): $e^- e^+  \to \gamma^* \gamma \to B \bar{B}  \gamma$ and its amplitude  is expressed in terms of the timelike baryon EM FFs.

\section{Generalized distribution amplitudes  and timelike form factors }
\label{ffs}

For a spin-$\tfrac{1}{2}$ baryon-antibaryon pair, the timelike  FFs of vector and axial-vector currents are defined as
\begin{equation}
\begin{aligned}
 \langle \bar{B}(p_2) B(p_1)  | \,\bar{q}(0) \gamma^{\mu} q(0)\, | 0 \rangle & = F_V^q(\hat{s}) \bar{u}(p_1) \gamma^{\mu} v(p_2) +F_S^q(\hat{s}) \frac{\Delta^{\mu}}{2m} \bar{u}(p_1)  v(p_2), \\
\langle \bar{B}(p_2) B(p_1)  | \,\bar{q}(0) \gamma^{\mu} \gamma_5 q(0)\, | 0 \rangle &= F_A^q(\hat{s}) \bar{u}(p_1) \gamma^{\mu} \gamma_5 v(p_2) +F_P^q(\hat{s}) \frac{P^{\mu}}{2m} \bar{u}(p_1) \gamma_5 v(p_2),
\label{eqn:emff}
\end{aligned}
\end{equation}
and a linear combination of $F_V^q(\hat{s})$ and $F_S^q(\hat{s})$ leads to the timelike electric and magnetic FFs,
\begin{equation}
\begin{aligned}
G_E(\hat{s})& =F_V^q(\hat{s}) + (\tau-1)F_S^q(\hat{s}), \\
G_M(\hat{s})&=F_V^q(\hat{s}),
\label{eqn:emff12}
\end{aligned}
\end{equation}
where $\tau=\hat{s}/(4m^2)$.
These timelike FFs are related to  the first moments of baryon  GDAs~\cite{Diehl:2002yh}
\begin{equation}
\begin{aligned}
\int_0^1 dz \Phi_i^q(z,\zeta_0, \hat{s}) &=F_i^q(\hat{s})  \, \, \text{for} \, \, i=V,\, A,\, P, \\
\int_0^1 dz \Phi_S^q(z,\zeta_0, \hat{s}) &=  \zeta_0 F_S^q(\hat{s}),
\label{eqn:emffgda}
\end{aligned}
\end{equation}

\begin{figure}[ht]
\centering
\includegraphics[width=0.8\textwidth]{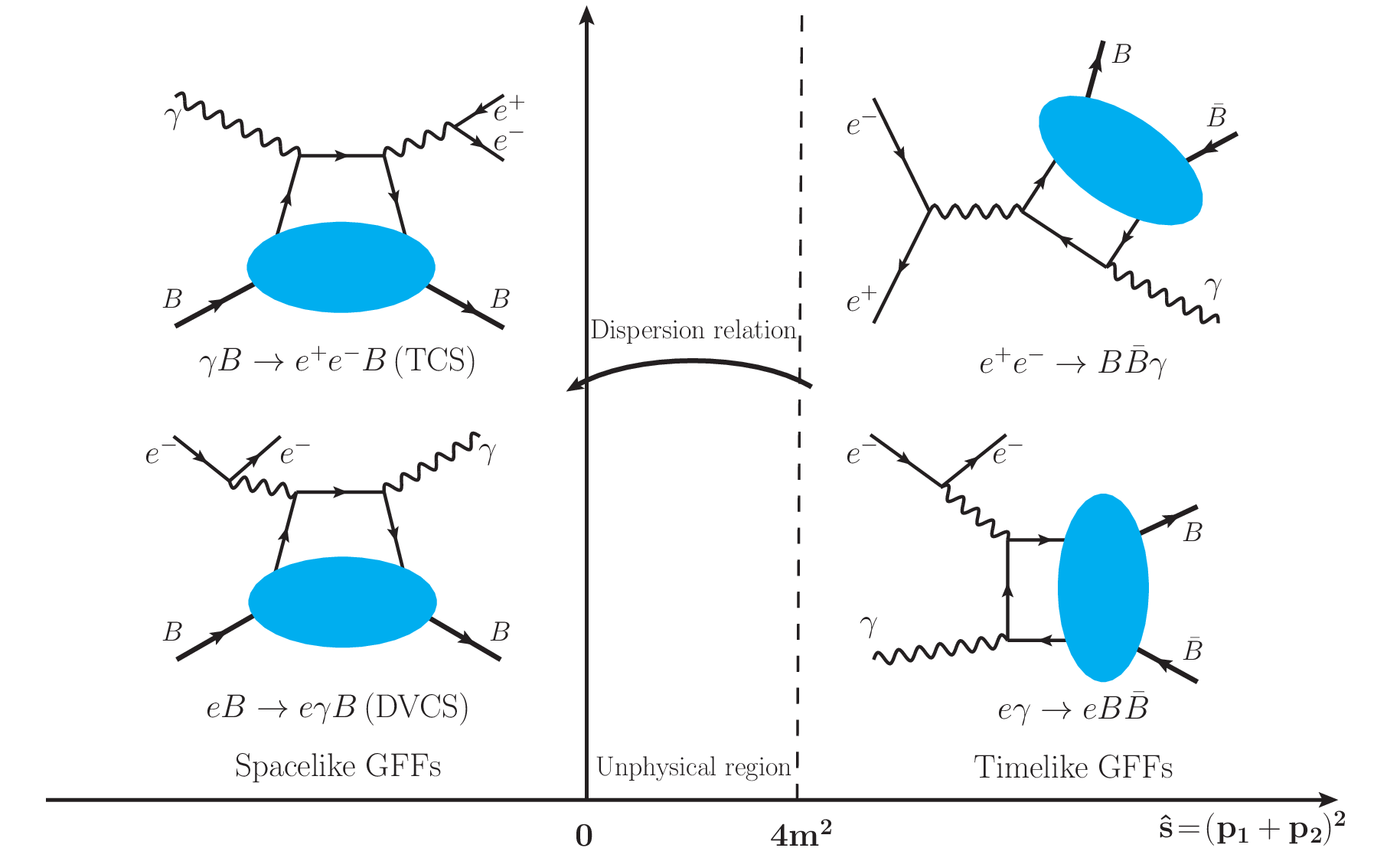}
\caption{ Spacelike and timelike EMT FFs are probed by different reactions, and  the former ones can be obtained from the latter ones using  dispersion relation.}
\label{fig:dr}
\end{figure}

The hadronic matrix elements of the EMT operator are parameterized in terms of  the EMT FFs. In the spacelike region, one finds the parametrization of these matrix elements in Refs.~\cite{Pagels:1966zza,Kobzarev:1962wt,Diehl:2003ny, Ji:1996ek, Bakker:2004ib, Lorce:2022cle}.
The timelike EMT FFs (GFFs) for a baryon  can be obtained by taking the $s$-$t$ crossed expression of the corresponding spacelike matrix element,
\begin{equation}
\begin{aligned}
& \langle \bar{B}(p_2) B(p_1)  | \,T_q^{\mu \nu}(0)\, | 0 \rangle \\
=&   \bar{u}(p_1) \left [ A^q(\hat{s}) \frac{\Delta^{\mu}\Delta^{\nu} }{4m}  +D^q(\hat{s}) \frac{P^{\mu}P^{\nu}-g^{\mu \nu} \hat{s} }{4m} -J^q(\hat{s}) \frac{\Delta^{\{ \mu} i \sigma^{\nu \} P} }{2m}   +\bar{C}^q(\hat{s}) m g^{\mu \nu}   \right]v(p_2).
\label{eqn:emt}
\end{aligned}
\end{equation}
One can use dispersion relations to relate
the timelike GFFs to the spacelike ones, as illustrated in Fig.~\ref{fig:dr}.  Thus, we can acquire a complete picture of the EMT FFs through the study of the timelike ones.
The timelike GFF $\bar{C}^q(\hat{s})$ of
Eq.~\eqref{eqn:emt} breaks the conservation law of EMT, and it must vanish when summing over quark flavors and gluon.
These EMT FFs are expressed by the second moments of baryon GDAs except for $\bar{C}^q(\hat{s})$,
\begin{equation}
\begin{aligned}
\int_0^1 dz (2z-1) \Phi_V^q(z,\zeta_0, \hat{s})  &= -2 \zeta_0 J^q(\hat{s}),  \\
\int_0^1 dz (2z-1) \Phi_S^q(z,\zeta_0, \hat{s})&=D^q(\hat{s})+  \left[ A^q(\hat{s})-2 J^q(\hat{s}) \right ] (\zeta_0)^2,
\label{eqn:emtsum}
\end{aligned}
\end{equation}

The inner structure of hadrons can be revealed with the help of the EMT FFs. For a spin-$\tfrac{1}{2}$ baryon, the angular momentum (AM) carried by quark flavor $q$ can be given by the EMT FF  $J^q(0)=J_z^q$, which is the sum of the orbital AM $L_z^q$ and helicity contribution $S_z^q$. The latter is related to the axial FF $F_A^q(\hat{s})$ of Eq.~\eqref{eqn:emff} as $S_z^q=\tfrac{1}{2}F_A^q(0)$, and the quark orbital AM is expressed as~\cite{Ji:1996ek}
\begin{align}
L_z^q=J^q(0)-\frac{1}{2}F_A^q(0),
\label{eqn:sl}
\end{align}
Thus, the spin decomposition of the baryons, nucleons as well as $\Lambda$, $\Sigma$  and $\Xi$, can be probed by studying the leading-twist GDAs of Eq.~\eqref{eqn:gdame}.
In addition, we can use the EMT FF $D^q(\hat{s})$ to study the $D$-terms of  baryons~\cite{Polyakov:1999gs},
\begin{align}
D=\sum_q D^q(0)+D^g(0),
\label{eqn:dterm}
\end{align}
where  $D^g$ is the gluonic EMT FF, and is related to the gluonic GPDs or GDAs. The $D$-term is a renormalization-scale invariant quantity, and it is also known as the last global unknown charge of hadrons, which has attracted a lot of attention in the past decades.

\section{Scattering amplitudes and cross sections }
\label{crs}
To describe the leading twist amplitude of subprocess (a) in Fig.~\ref{fig:gda}, we first introduce the  hadron tensor for $\gamma^* \to B \bar{B}  \gamma$,
\begin{align}
T_{\mu \nu}=i\int d^4x\, e^{-iq_1\cdot x} \langle \bar{B}(p_2) B(p_1)  | \,
T \{ j_{\mu}^{\text{em}}(x)  j_\nu^{\text{em}} (0) \} \, | 0 \rangle,
\label{eqn:amp0}
\end{align}
where a $C$-even baryon-antibaryon pair is produced.
At leading twist, the  hadron tensor $T_{\mu \nu}$ only receives  contribution from the transversely polarized photons,
\begin{align}
T^{\mu \nu}=\frac{-2 }{ \sqrt{2s}} \left \{ g_{T }^{ \mu \nu }\left [ \zeta_0 \mathcal{F}_{V}  \bar{u}  \! \left ( p_{1}  \right )\! \gamma ^{+}  \! v\!\left ( p_{2}  \right )  + \mathcal{F}_{S}  \frac{P^{+} }{ 2m } \bar{u}\!\left ( p_{1}  \right )\! v\!\left ( p_{2}  \right )     \right ] -i\epsilon  _{T}^{\mu \nu} \left [ \mathcal{F}_{A}  \bar{u} \! \left ( p_{1}  \right )\! \gamma ^{+}  \!\gamma ^{5}\! v\!\left ( p_{2}  \right ) +\mathcal{F}_{P}  \frac{P^{+} }{2m } \bar{u}\!\left ( p_{1}  \right )\! \gamma ^{5} \! v\!\left ( p_{2}  \right )   \right ]   \right \},
\label{eqn:tmunu}
\end{align}
where the transverse tensor  $\epsilon_{T}^{\mu \nu} = \epsilon^{\mu \nu \alpha \beta} \tilde{n}_{\alpha} n_{\beta}$ is defined,  and the  hadron tensor satisfies the electromagnetic gauge invariance condition $T_{\mu \nu}q_1^{\mu}=T_{\mu \nu}q_2^{\nu}=0$. The functions $\mathcal{F}_{i}$ can be regarded as the timelike version of the Compton FFs in DVCS, and they are expressed at leading (zeroth) order in $\alpha_s$ in term of the four leading-twist GDAs,
\begin{equation}
\begin{aligned}
     \zeta_0 \mathcal{F}_{V} =&\sum_{q} \frac{e_{q}^{2}}{2}\int_{0}^{1} dz\frac{2z-1}{ z\bar{z}  }\Phi _{V}^{q }\!\left ( z,\zeta_0 ,\hat{s}  \right ),  \\
     \mathcal{F}_{S} =&\sum_{q} \frac{e_{q}^{2}}{2}\int_{0}^{1} dz\frac{2z-1}{ z\bar{z}  }\Phi _{S}^{q }\!\left ( z,\zeta_0 ,\hat{s}  \right ),  \\
     \mathcal{F}_{i} =&\sum_{q} \frac{e_{q}^{2}}{2}\int_{0}^{1} dz\frac{1}{ z \bar{z}   }\Phi _{i}^{q }\!\left ( z,\zeta_0,\hat{s}  \right ) \quad \text{for} \quad i=A,\, P,
\label{cffs}
\end{aligned}
\end{equation}
where only the $C$-even GDAs contribute, and they are defined as
\begin{equation}
\begin{aligned}
      \Phi _{i}^{q+}\!\left ( z,\zeta_0,\hat{s}  \right ) =& \tfrac{1}{2} \left[ \Phi _{i}^{q }\!\left ( z,\zeta_0,\hat{s}  \right ) - \Phi _{i}^{q }\!\left (
     \bar{z},\zeta_0,\hat{s}  \right ) \right] \quad \text{for} \quad i=V,\, S, \\
     \Phi _{i}^{q+}\!\left ( z,\zeta_0,\hat{s}  \right ) =& \tfrac{1}{2} \left[ \Phi _{i}^{q }\!\left ( z,\zeta_0,\hat{s}  \right ) + \Phi _{i}^{q }\!\left (
     \bar{z},\zeta_0,\hat{s}  \right ) \right] \quad \text{for} \quad i=A,\, P.
\end{aligned}
\end{equation}
Then, the contribution of subprocess (a) to the differential cross section  is expressed as
\begin{equation}
\begin{aligned}
  \frac{d\sigma _{\mathrm{G} } }{d\hat{s}dud\left ( \cos \theta  \right ) d\varphi    } = & \frac{\alpha _{\mathrm{em} }^{3} \beta _{0}}{8\pi s^{3}   }\frac{1}{1+\epsilon }  \Bigg [   \left  | \mathcal{F} _{A}  \right |^{2} -\left | \mathcal{F} _{S}  \right | ^{2}+2\mathrm{Re} \left ( \mathcal{F} _{A}\mathcal{F} _{P}^{\ast } \right)+\frac{\hat{s} \left ( \left | \mathcal{F} _{P}  \right |^{2} +\left | \mathcal{F} _{S}  \right | ^{2}   \right ) }{4m^{2}  }  \\
  & + (\beta _{0})^2 \cos^2\theta \left[  \left | \mathcal{F} _{V}  \right |^{2} +  2\mathrm{Re} \left ( \mathcal{F} _{S}\mathcal{F} _{V}^{\ast } \right )- \left | \mathcal{F} _{A}  \right | ^{2} \right]
   -(\beta _{0})^4 \cos^4\theta \left | \mathcal{F} _{V}  \right |^{2}   \Bigg ],
   \label{crogda}
   \end{aligned}
\end{equation}
where the $\epsilon$  parameter  is given by
\begin{align}
\epsilon=\frac{y-1}{1-y+\frac{y^2}{2}}, \qquad y=\frac{q_1\cdot q_2}{k_1\cdot q_2}.
\label{eqn:polar}
\end{align}

The contribution of the ISR subprocess (b) to the cross section  has been investigated in Refs.~\cite{Czyz:2007wi, Faldt:2013gka}, and depends on the baryon EM FFs of Eqs.~\eqref{eqn:emff} and \eqref{eqn:emff12}. In this work, this cross section will be discussed in the center-of-mass frame of the baryon-antibaryon pair, and it is given by
\begin{equation}
     \frac{d\sigma _{\mathrm{ISR} } }{d\hat{s}dud\left ( \cos \theta  \right ) d\varphi    }=\frac{\alpha _{\mathrm{em} }^{3}\beta_{0}^{3}   }{4\pi s^{2} }\frac{1}{\epsilon \hat{s} }  \left [ b_{0}+b_{1} \cos ^{2} \theta +b_{2}  \sin ^{2}\theta +b_{3} \sin \left ( 2\theta  \right )\cos \varphi +b_{4} \sin ^{2}\theta \cos \left ( 2\varphi  \right )   \right ],
\label{eqn:isrcro}
\end{equation}
where the coefficients $b_i$ read
\begin{equation}
\begin{aligned}
b_{0}&=\left [ 1-2x\left ( 1-x \right )  \left ( 1+\epsilon  \right )  \right ] (2\lambda-1)  \left | G_{M}  \right |^{2}   ,\\
b_{1}&=\left [ 1-2x \left ( 1-x \right )\left ( 1-\epsilon  \right )   \right ]  \left | G_{M} \right |^{2}+4\epsilon x\left ( x-1 \right ) (\lambda-1) \left[ \left | G_{E} \right |^{2} -\left |G_{M}\right |^{2} \right] ,\\
b_{2}&=2\epsilon x\left ( x-1 \right ) \left | G_{M}\right |^{2} +\left [ 1-2x\left ( 1-x \right ) \right ] (\lambda-1) \left[ \left | G_{E} \right |^{2} -\left |G_{M}\right |^{2} \right]  ,\\
b_{3}&=\sqrt{\epsilon \left ( 1-\epsilon  \right ) }\sqrt{2x\left ( x-1 \right ) }\left ( 2x-1 \right ) \text{sgn}\!\left ( \rho  \right )\left[  (\lambda-1) \left |G_{E} \right |^{2} -\lambda \left |G_{M}\right |^{2} \right], \\
b_{4}&=2\epsilon  x\left ( 1-x \right )  \left[  (\lambda-1) \left |G_{E} \right |^{2} -\lambda \left |G_{M}\right |^{2} \right].
\label{eqn:coisr}
\end{aligned}
\end{equation}
The dimensionless parameters $x$ and $\lambda$ are defined as $x=s/(s-\hat{s})$ and $\lambda=1/(\beta_0)^2$, respectively. In Eq.~\eqref{eqn:coisr}, $\text{sgn}(\rho)$ is the sign function for the variable $\rho=\hat{s}-s-2u$.

In addition, we need to consider the  contribution coming from the interference of the two amplitudes for subprocesses (a) and (b),
\begin{equation}
\frac{d\sigma _{\mathrm{I} } }{d\hat{s}dud\left ( \cos \theta  \right ) d\varphi  } =\frac{\alpha  _{\mathrm{em} }^{3} \beta _{0}  }{8\pi s^{2} }\frac{\sqrt{2}\beta _{0}  }{\sqrt{\hat{s} s\epsilon (1+\epsilon )} } \left [ c_{0} \cos \theta +c_{1} \cos ^{3}\theta +c_{2} \sin \theta \cos \varphi +c_{3} \sin \left ( 2\theta  \right ) \cos \theta \cos \varphi  \right ],
\label{intcross}
\end{equation}
where the coefficients $c_i$ are expressed as
\begin{equation}
\begin{aligned}
c_{0}=&2\text{sgn}\left ( \rho  \right ) \sqrt{\epsilon \left ( 1-\epsilon  \right ) }  \sqrt{2x\left ( x-1 \right )  } \left[  \mathrm{Re} \left ( \mathcal{F}_{V} G_{M}^{\ast } \right ) +\mathrm{Re}(\mathcal{F}_{S} G_{E}^{\ast }) \right] ,\\
c_{1}=&2(\beta _{0})^2\text{sgn}\left ( \rho  \right )\sqrt{\epsilon \left ( 1-\epsilon  \right ) }  \sqrt{2x\left ( x-1 \right )  }
\left[ (\lambda-1) \mathrm{Re}\left( \mathcal{F}_{V}G_{E}^{\ast }\right ) -\lambda \mathrm{Re}(\mathcal{F}_{V}G_{M}^{\ast } ) \right]  ,\\
c_{2}=&2\left [ 1-\left ( 1-x \right )  \left ( 1+\epsilon  \right )  \right ] \mathrm{Re}\left ( \mathcal{F} _{A}G_{M}^{\ast }   \right ) +2 \left [ 1-\left ( 1-x \right ) \left ( 1-\epsilon  \right )  \right ]
 \mathrm{Re}(\mathcal{F}_{S} G_{E}^{\ast }),\\
c_{3}=&(\beta _{0})^2 \left [ 1-\left ( 1-x \right ) \left ( 1-\epsilon  \right )  \right ]\left[ (\lambda-1) \mathrm{Re}\left( \mathcal{F}_{V}G_{E}^{\ast }\right ) -\lambda \mathrm{Re}(\mathcal{F}_{V}G_{M}^{\ast } ) \right].
\label{intcros}
\end{aligned}
\end{equation}

The timelike Compton FFs depend on the baryon GDAs, which only appear in the cross sections of Eqs.~\eqref{crogda} and \eqref{intcross}. There are no Compton FFs in the cross section of the ISR subprocess (b), which  gives the largest contribution in  $e^- e^+  \to B \bar{B}  \gamma$; this may be understood,
provided that the timelike Compton FFs and GFFs are of the comparable magnitude, by  the simple scaling rule :
\begin{align}
d \sigma_{\text{G}} :d \sigma_{\text{I}}: d \sigma_{\text{ISR}} \sim  \frac{1}{s}:\frac{1}{\sqrt {s \hat s}} : \frac{1}{\hat s}.
\label{eqn:cro-3}
\end{align}
A full extraction of baryon GDAs from the cross section necessitates studying the interference contribution in Eq.~\eqref{intcross}. This is because baryon GDAs are complex functions, meaning Eq.~\eqref{crogda} alone cannot fully constrain the reconstruction of the timelike Compton FFs. Crucially, the interference term allows the extraction of the Compton FFs' imaginary phases, provided the EM FFs are used as inputs.
If we consider the exchange of $(\theta, \varphi) \to (\pi-\theta, \pi+\varphi)$ in the cross sections,  Eqs.~\eqref{crogda} and \eqref{eqn:isrcro} will remain the same, however, a minus sign will appear in Eq.~\eqref{intcross} because of the different charge conjugation of the baryon-antibaryon pair in the subprocesses (a) and (b).
Thus, the interference contribution can be obtained from taking the difference of the cross-section
in this exchange, $d \sigma(B\bar{B})-d \sigma(\bar{B}B)=2d \sigma_{\text{I}} $.

\section{Numerical estimates of  the cross sections}
\label{nucr}

For the numerical calculation of the cross sections, we need to use the EM FFs and GDAs as inputs, and we explain here the models of the EM FFs and GDAs that are adopted in this work. The effective proton EM FF \footnote{There exist other parametrizations \cite{Lomon:2012pn} which incorporate both the vector meson dominance model and analytic continuation techniques to develop a unified description of timelike and spacelike EM FFs for nucleons. Using such a parametrization will not make much difference in our analysis since experimental data constrain all models in the kinematical range we explore.} contains two ingredients~\cite{Bianconi:2015owa},
\begin{align}
F_p(\hat{s})=F_{\rm 3p}(\hat{s})\ +\ F_{\rm osc}(p),
\label{eq:diff}
\end{align}
where the three-pole component $F_{\rm 3p}(s)$ gives the dominant contribution and  $F_{\rm osc}(p)$ describes a periodic oscillating behavior.  These functions are expressed as
\begin{equation}
\begin{aligned}
F_{\rm 3p}(\hat{s})&= \frac{F_{0}} {\left(1+\frac{\hat{s}}{m_a^2}\right)\left(1-\frac{\hat{s}}{m_0^2} \right)^2}, \\
F_{\rm osc}(p)&=Ae^{-Bp}\cos(Cp+D),
\label{eq:fosc}
\end{aligned}
\end{equation}
with
\begin{align}
p=\sqrt{\hat{s}\left(\frac{\hat{s}}{4m^2}-1\right)}.
\label{eq:p-mom}
\end{align}
The modulus of the ratio of  the timelike electric and magnetic FFs is denoted as $R(\hat{s})=|G_E(\hat{s})|/|G_M(\hat{s})|$, and is parameterized as~\cite{Tomasi-Gustafsson:2020vae}
\begin{align}
R(\hat{s})=
\displaystyle\frac{1}{1+\omega^2 /r_0} \left [1+r_1e^{-r_2 \omega }\sin\left(r_3 \omega
\right)\right ], \, \, \,  \omega=\sqrt{\hat{s}}-2m.
\label{eq:FR}
\end{align}
Then, $G_E$ and $G_M$ are expressed in terms of the effective EM FF $F_p$ and the ratio $R$,
\begin{equation}
\begin{aligned}
|G_E(\hat{s})|&=F_p(\hat{s})\sqrt{ \frac{1+2\tau}{1+2\tau/R(\hat{s})^2}}, \\
|G_M(\hat{s})|&=F_p(\hat{s})\sqrt{ \displaystyle\frac{1+2\tau}{R(\hat{s})^2+2\tau}}.
\label{eq:gegm}
\end{aligned}
\end{equation}
The authors of Ref.~\cite{Tomasi-Gustafsson:2020vae} perform a global analysis of recent experimental measurements on $e^+e^- \to p \bar{p}$, and the  parameters of the effective FF $F_p$ and the ratio $R$ are summarised in Table \ref{Tab:Rfit}. We neglect the imaginary phases of EM FFs $G_E$ and $G_M$ in the numerical estimate, as our knowledge of the imaginary phases is quite limited at the current stage.

\begin{table}[ht]
 \caption{\label{Tab:Rfit} Parameters are obtained from a global fit  for the effective FF $F_p$ and the ratio $R$~\cite{Tomasi-Gustafsson:2020vae}.}
\centering
\begin{tabular}{|c|c|c|c|c|c|}
\hline
$F_0$  & 9.7  &$D$ & \,\,\,0.04\,\,\, & $r_1$ & 0.5 \\
\hline
$A$ & \,0.073 \, & $m_0^2$ (GeV$^2$) & 0.71 & $r_2$ (GeV$^{-1}$)& \, \,1.5 \, \, \\
\hline
$B$ (GeV$^{-1}$)& 1.05 &  $m_a^2$ (GeV$^2$) & 7.1 &$r_3$ (GeV$^{-1}$)& 9.3\\
\hline
$C$ (GeV$^{-1}$)& 5.51& $r_0$ & 3 &  & \\
\hline
\end{tabular}
\end{table}

The proton-antiproton GDAs  or the timelike Compton FFs are completely unknown.
In this work, we assume that the vector GDAs are of a  similar magnitude for the proton-antiproton and dipion meson cases,
\begin{align}
   \Phi _{S}^{q } \sim  \Phi _{V}^{q } \sim  \Phi _{\pi \pi}^{q },
 \label{cffpion}
\end{align}
and the latter has been extracted from experimental measurements~\cite{Kumano:2017lhr}. The pion vector GDA is expressed as
\begin{align}
\Phi^q_{\pi\pi }(z, \cos\theta,  \hat{s})  =
 \frac{3 (2 \alpha +3 )}{5B(\alpha+1,\alpha+1)} z^\alpha(1-z)^\alpha (2z-1)
\left [ \widetilde{B}_{10}^q(\hat{s})+\widetilde{B}_{12}^q(\hat{s})P_2(\cos\theta) \right],
\label{eqn:phi-paramet1}
\end{align}
where $\alpha=1.157$, and $B(a,b)$ is Euler's beta function. The complex functions
$\widetilde{B}_{nl}$ are dependent on the invariant mass of the pion pair, and
their imaginary parts are given by the phases $\delta_l$,
\begin{align}
\widetilde{B}_{nl}^q(\hat{s})=\bar{B}_{nl}^q(\hat{s}) \, e^{i\delta_l} .
\label{eqn:b_phase}
\end{align}
Since  the relative phase $\delta_0-\delta_2$ is only determined below $\hat{s}=4$ GeV$^2$ through the analysis of Belle measurements, and since perturbative QCD arguments show that  the phases will be suppressed by $\alpha_s$ for the large  $\hat{s}$-region~\cite{Diehl:1999ek, Song:2025zwl}, we will ignore the phases $\delta_l$ in our numerical estimate of the proton pair production.
The real parts of  $\widetilde{B}_{nl}$ are parameterized as
\begin{equation}
\begin{aligned}
\bar{B}_{10}^q(\hat{s})& = - \frac{10R_{q}}{9}\left( 1 +\frac{2m^2}{\hat{s}} \right) \, F_q(\hat{s}), \\
\bar{B}_{12}^q(\hat{s})& = \frac{10R_{q}}{9} \left( 1 -\frac{4m^2}{\hat{s}} \right) \, F_q (\hat{s}),
\label{eqn:res}
\end{aligned}
\end{equation}
with
\begin{align}
 F_q (\hat{s}) = \frac{1}{\left[ 1 + (\hat{s}-4 m^2)/\Lambda^2 \right]},
\end{align}
where $R_u=1/3$, $R_d=1/6$, $\Lambda=1.928$ GeV, and we replace the pion mass with the proton mass $m$.
The resonance contribution is neglected in Eq.~\eqref{eqn:res}. The corresponding Compton FF is expressed as
\begin{align}
 \mathcal{F}_{\pi} =\sum_{q} \frac{e_{q}^{2}}{2}\int_{0}^{1} dz\frac{2z-1}{ z\bar{z}  }\Phi _{\pi \pi}^{q }\!\left ( z,\zeta_0 ,\hat{s}  \right ).
\label{cffpi}
\end{align}
Combining Eqs.~\eqref{cffpi}, \eqref{cffpion}, and \eqref{cffs}, we can derive
\begin{equation}
\begin{aligned}
\mathcal{F}_{S} & = \mathcal{F}_{\pi}, \\
\zeta_0 \mathcal{F}_{V}  & =\mathcal{F}_{\pi}\cos \theta ,
\label{eqn:cffpionp}
\end{aligned}
\end{equation}
where $\cos \theta$ appears to  make sure that $\mathcal{F}_{V}$ is a even function of $\cos \theta$.
Since there are no studies on the axial-vector GDAs  $\Phi _{S}^{q }$ and  $\Phi _{V}^{q }$, we just express their corresponding  Compton FFs as
\begin{align}
\mathcal{F}_{A} =\mathcal{F}_{P} =g_A \mathcal{F}_{\pi},
\label{eqn:sdep-ff1}
\end{align}
where  $g_A=e_u g_A^u +e_d g_A^d$, and the recent lattice QCD results indicate  $g_A^u=0.832$ and $g_A^d=-0.417$ for the axial charges~\cite{Alexandrou:2024ozj}.

Now we use the models of EM FFs and GDAs we have discussed to calculate the cross sections for the process $ e^+ e^- \to p \bar{p}  \gamma$. After integration over azimuthal angle, the differential cross sections are expressed as
\begin{equation}
\begin{aligned}
 \frac{d\sigma _{\mathrm{ISR} } }{d\hat{s}dud\left ( \cos \theta  \right )     }=&\frac{\alpha _{\mathrm{em} }^{3}\beta_{0}^{3}   }{2s^{2} }\frac{1}{\epsilon \hat{s} }  \left [ b_{0}+b_{1} \cos ^{2} \theta +b_{2}  \sin ^{2}\theta   \right ],\\
\frac{d\sigma _{\mathrm{G} } }{d\hat{s}dud\left ( \cos \theta  \right )    } = & \frac{\alpha _{\mathrm{em} }^{3} \beta _{0}}{4 s^{3}   }\frac{1}{1+\epsilon }  \Bigg [   \left  | \mathcal{F} _{A}  \right |^{2} -\left | \mathcal{F} _{S}  \right | ^{2}+2\mathrm{Re} \left ( \mathcal{F} _{A}\mathcal{F} _{P}^{\ast } \right)+\frac{\hat{s} \left ( \left | \mathcal{F} _{P}  \right |^{2} +\left | \mathcal{F} _{S}  \right | ^{2}   \right ) }{4m^{2}  }  \\
  & + (\beta _{0})^2 \cos^2\theta \left[  \left | \mathcal{F} _{V}  \right |^{2} +  2\mathrm{Re} \left ( \mathcal{F} _{S}\mathcal{F} _{V}^{\ast } \right )- \left | \mathcal{F} _{A}  \right | ^{2} \right]
   -(\beta _{0})^4 \cos^4\theta \left | \mathcal{F} _{V}  \right |^{2}   \Bigg ], \\
 \frac{d\sigma _{\mathrm{I} } }{d\hat{s}dud\left ( \cos \theta  \right )    }=&\frac{\alpha _{\mathrm{em} }^{3}(\beta_{0})^{2}   }{4 s^{2} }\frac{\sqrt{2}}{\sqrt{s\hat{s} \epsilon(1+\epsilon) } }  \left [ c_{0} \cos \theta +c_{1} \cos ^{3} \theta  \right ].
\label{eqn:croint}
\end{aligned}
\end{equation}
The maximum values  are $\sqrt{s}=5.6$ GeV and $\sqrt{s}=7$ GeV for the center-of-mass energy of $e^+e^-$ at  BESIII and  the  proposed  STCF, respectively, and the designed peak luminosity of STCF can be $0.5 \times 10^{35}$ cm$^{-2}$s$^{-1}$ or higher~\cite{Achasov:2023gey}, which indicates that precise measurements of $e^- e^+ \to B \bar{B}  \gamma$ will be  possible at STCF. We choose $\sqrt{s}=4$ GeV  in this numerical estimate, and this value should be typical for these two experimental facilities. To satisfy the QCD factorization condition $s \gg \hat{s}$, we use the range  $2.0$ GeV $<W=\sqrt{\hat{s}} <2.7$ GeV for the invariant mass of the proton-antiproton pair.

\begin{figure}[htb]
\centering
\includegraphics[width=0.6\textwidth]{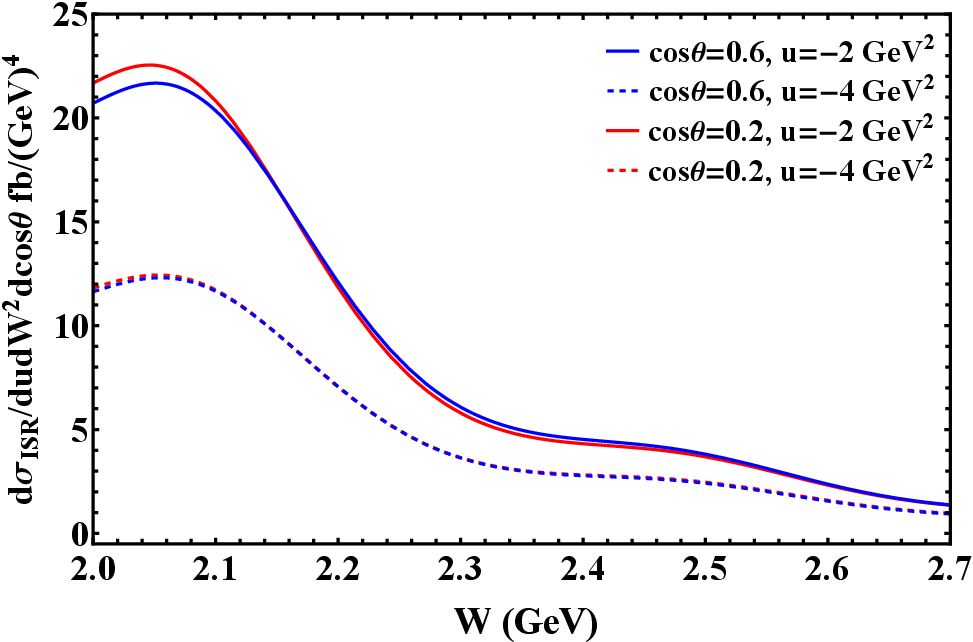}
\caption{ The differential cross section of the ISR process $e^- e^+  \to \gamma^*  \gamma \to p \bar{p}  \gamma$.}
\label{fig:crossisr}
\end{figure}

\begin{figure}[htb]
\centering
\includegraphics[width=0.6\textwidth]{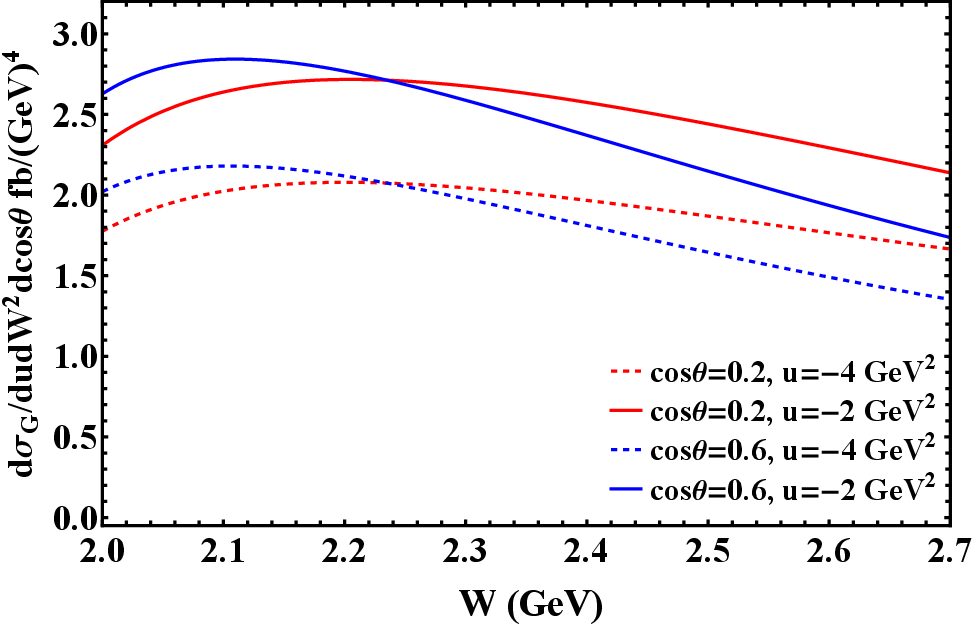}
\caption{ The differential cross section of $e^- e^+  \to \gamma^*  \to p \bar{p}  \gamma$.}
\label{fig:crossg}
\end{figure}

The cross section of the ISR process is depicted  in Fig.~\ref{fig:crossisr},
and we choose $\cos\theta =0.2$ and $\cos\theta =0.6$ for the red and blue lines, respectively.
The variable $u=-2$ GeV$^2$ is fixed for the solid lines, and  the dashed ones are for $u=-4$ GeV$^2$.
Since we only use the magnitude of baryon EM FFs $|G_E|$ and $|G_M|$ as inputs, and these FFs have been constrained in the recent experiments, the result for the ISR process should be regarded as a rather precise prediction.  Moreover, this process is not described in terms of a QCD twist expansion.
In Fig.~\ref{fig:crossg}, the differential  cross section of $e^- e^+  \to \gamma^*  \to p \bar{p}  \gamma$ is provided,
with same conventions   as those in Fig.~\ref{fig:crossisr}.
The ISR contribution is much larger than the GDA contribution as expected.
However, the latter does not decrease rapidly with the growing invariant mass of the baryon pair, which is different from the ISR case.
This  behavior appears due to the functional form choosen here for the nucleon-antinucleon GDAs.
As our knowledge is quite limited for the baryon GDAs, further theoretical and experimental studies on the baryon-antibaryon GDAs are necessary for a more precise evaluation of this GDA contribution.

\section{Baryon-antibaryon asymmetry}
\label{Asymm}
\begin{figure}[htb]
\centering
\includegraphics[width=0.90\textwidth]{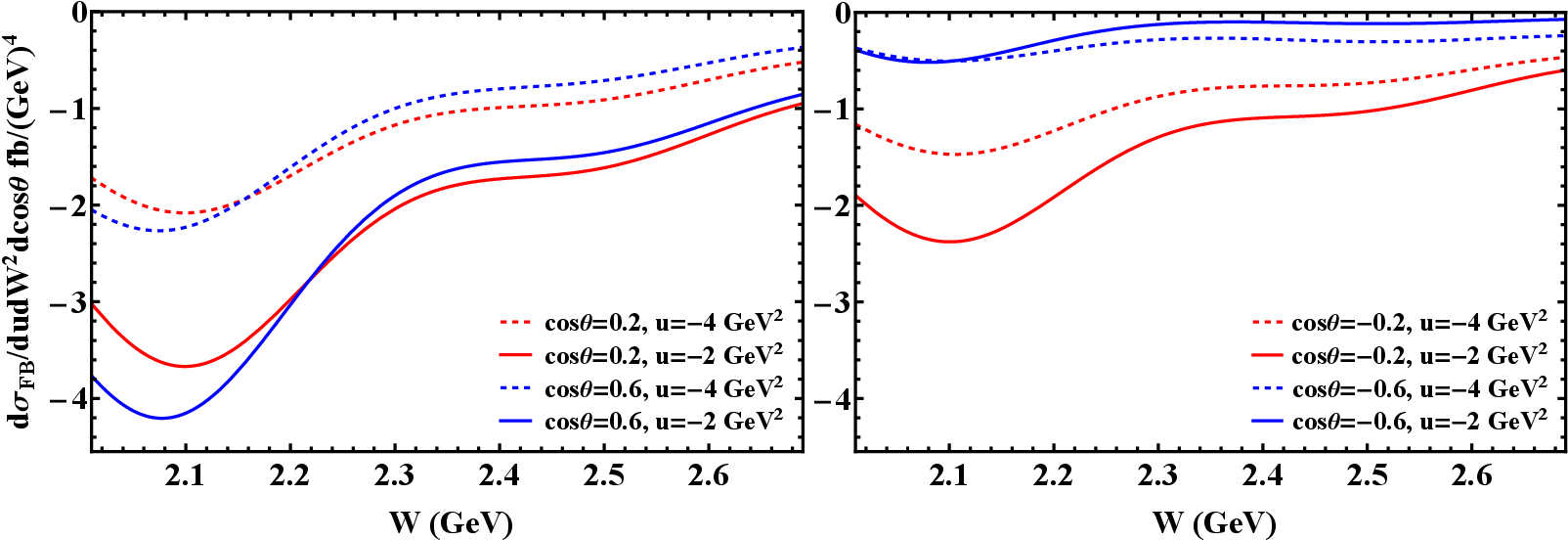}
\caption{ The cross section difference (Eq.\ref{intasym}) allows to measure the interference term of $e^- e^+  \to p \bar{p}  \gamma$. It depends on the invariant mass of the proton-antiproton pair and on the polar angle $\theta$ of the nucleon.}
\label{fig:crossint}
\end{figure}

The different charge-conjugation properties of the ISR and QCD process allows to define and exploit a new and most interesting observable, namely a baryon-antibaryon asymmetry. This procedure is very similar to the charge asymmetry observable defined in the TCS process \cite{Berger:2001xd} and used in the first observation of the TCS process in the JLab-CLAS experiment \cite{CLAS:2021lky}.
Since in the $B \bar B$ center of mass frame, the baryon and antibaryon $3-$momenta are opposite, one may define a differential asymmetry as
\begin{equation}
    A (\theta,\varphi) =\frac{\frac{d\sigma (\theta,\varphi)}{d\cos \theta ~ d \varphi}- \frac{d\sigma (\pi -\theta,\varphi+\pi)}{d\cos \theta ~ d \varphi}}{\frac{d\sigma (\theta,\varphi)}{d\cos \theta ~ d \varphi}+ \frac{d\sigma (\pi -\theta,\varphi+\pi)}{d\cos \theta ~ d \varphi}},
\end{equation}
or one can integrate over one hemisphere (i.e. $\pi/2 < \varphi < 3\pi/2$) and subtract the integration over the complementary region, to get a forward-backward asymmetry:
\begin{equation}
    A_{FB} (\theta) =\frac{\int_{\pi/2}^{3\pi/2} d\varphi\frac{d\sigma (\theta,\varphi)}{d\cos \theta ~ d \varphi}- \int_{3\pi/2}^{2\pi} d \varphi\frac{d\sigma (\pi -\theta,\varphi)}{d\cos \theta ~ d \varphi}- \int_{0}^{\pi/2} d \varphi\frac{d\sigma (\pi -\theta,\varphi)}{d\cos \theta ~ d \varphi}}{\int_{\pi/2}^{3\pi/2} d\varphi\frac{d\sigma (\theta,\varphi)}{d\cos \theta ~ d \varphi}+ \int_{3\pi/2}^{2\pi} d \varphi\frac{d\sigma (\pi -\theta,\varphi)}{d\cos \theta ~ d \varphi}+ \int_{0}^{\pi/2} d \varphi\frac{d\sigma (\pi -\theta,\varphi)}{d\cos \theta ~ d \varphi}}\,.
\end{equation}
As one easily sees from Eq. \eqref{intcross}, the interference term is the only contribution that survives in the numerator and it does not contribute to the denominator. Denoting the numerator of $A_{FB}$ as $\int_{\Delta \varphi} d\varphi \frac{d\Delta\sigma (\theta,\varphi)}{d\cos \theta ~ d \varphi}$, one gets:
\begin{equation}
\begin{aligned}
 \frac{d\sigma _{\mathrm{FB} } }{d\hat{s}dud\left ( \cos \theta  \right )     }=&\int_{\Delta\varphi} d\varphi\frac{d\Delta\sigma (\theta,\varphi)}{d\cos \theta ~ d \varphi}   \\
 = &\frac{\alpha  _{\mathrm{em} }^{3} \beta _{0}  }{8\pi s^{2} }\frac{\sqrt{2}\beta _{0}  }{\sqrt{\hat{s} s\epsilon (1+\epsilon )} } \left [ 2\pi ( c_{0} \cos \theta + c_{1} \cos ^{3}\theta) -4 (c_{2} \sin \theta  + c_{3} \sin \left ( 2\theta  \right ) \cos \theta )  \right ].
\label{intasym}
\end{aligned}
\end{equation}
We show on Fig. \ref{fig:crossint} this difference of cross sections as
a function of  the invariant mass of the proton-antiproton pair $W$,  and $u=-2$ GeV$^2$ ($-4$ GeV$^2$) is chosen for the solid (dashed) lines. The colors (red, blue) of lines represent different values of $\cos\theta$ ($\pm0.2$, $\pm0.6$).
 The cross section increases from $W = 2.0$ GeV to $2.1$ GeV with the growing phase space, whereas beyond the peak it decreases as $W$ rises, due to diminishing Compton FFs and EM FFs.
We find that the interference term is larger than the pure QCD contribution in Fig.~\ref{fig:crossg}, which means that this baryon-antibaryon asymmetry should play an important role in the future extraction of baryon GDAs.
Let us stress that almost nothing is known about the baryon-antibaryon GDAs at the current stage, and the actual GDAs could be significantly different from the model we use in this work. Thus, our numerical results of  the  interference term and the GDA contribution can be only considered as  order-of-magnitude estimates for the cross section.

\section{Summary}
\label{summary}

The hadronic GPDs are three-dimensional functions of hadrons, which are accessed in the exclusive processes such as the DVCS reaction $\gamma^{*} h \to \gamma h$. The second moments of GPDs lead to the hadronic matrix elements of the EMT operator, which are parametrized in terms of EMT FFs. If we take the $s$-$t$ crossed channel of DVCS, namely the reaction  $\gamma^{*} \gamma  \to  h \bar{h}$,  hadronic GDAs are involved. GDAs can be regarded as the timelike crossed GPDs;
thus, the study of GDAs sheds light on the GPDs and vice versa.
The extraction of GDAs from experimental cross sections is possible even for the unstable hadrons, and corresponding measurements can be conducted at many electron-positron facilities.

In this work, we investigate the process $ e^+ e^- \to B \bar{B}  \gamma$, which depends on two subprocesses: the ISR process and $ e^+ e^- \to \gamma^* \to B \bar{B}  \gamma$.
The ISR process is described by the baryon EM FFs, and using QCD factorization, the amplitude of the latter subprocess can be factorized into two ingredients : the hard amplitude $\gamma^* \to q \bar{q}  \gamma$ and the non-perturbative hadronic matrix element describing the process
$q \bar{q} \to B \bar{B}$, known as the baryon-antibaryon GDAs. The relations between the baryon GDAs and timelike EMT FFs are discussed, and these FFs can be used to study the spin decompositions and  D-terms of the baryons. We calculate the cross sections of $ e^+ e^- \to B \bar{B}  \gamma$ at twist 2, and they are expressed in terms of the timelike baryon Compton FFs and EM FFs. Even though there are now no detailed theoretical or experimental studies on the baryon-antibaryon GDAs, we try to build a model for the proton-antiproton GDAs. For the proton EM FFs, they have been extracted from recent experimental measurements. Using these  proton-antiproton GDAs and EM FFs, we have provided numerical estimates for the  process $ e^+ e^- \to p \bar{p}  \gamma$, which should be helpful to  future measurements of this process at BESIII, Belle II, and the  proposed  STCF.

This work is but a first step in the analysis of the reaction $ e^+ e^- \to p \bar{p}  \gamma$. The mass of the nucleon makes quite uncertain the validity of the assumption $\hat s <<s$ where the GDA factorization is proven. Higher twist corrections of $O(\hat s/s)$ should be evaluated before any precise extraction of baryon-antibaryon GDAs. Part of these higher twist corrections, the so-called kinematic higher twist ones \cite{Braun:2011dg}, which have been estimated for DVCS and TCS \cite{Braun:2014sta, Martinez-Fernandez:2025gub}, can  readily be computed for this reaction, as it has been for the reaction $ e^+ e^- \to \pi \pi  \gamma$. \cite{Pire:2023kng, Pire:2023ztb}. We shall deal with these corrections, as well as with the leading-twist NLO corrections in $\alpha_s$, in a future work.

\section*{Acknowledgments}
We acknowledge useful discussions with Dr. Xiao-Yu Wang.  Qin-Tao Song was supported by the National Natural Science Foundation
of China under Grant Number 12005191.



\end{document}